\documentclass[preprint]{aastex}
\usepackage{graphicx}
\usepackage{amsmath}
\usepackage[caption=false]{subfig}
\usepackage{natbib}
\usepackage{booktabs}
\usepackage{array}
\newcolumntype{L}{>{\centering\arraybackslash}m{7cm}}

\begin{document}

\title{The development of a split-tail heliosphere and the role of non-ideal processes: a comparison of the BU and Moscow models}

\author{M. Kornbleuth\altaffilmark{1}}
\affil{\altaffilmark{1}Astronomy Department, Boston University, Boston, MA 
02215, USA} 
\email{kmarc@bu.edu}

\author{M. Opher\altaffilmark{1}$^{,}$\altaffilmark{2}}
\affil{\altaffilmark{2}Radcliffe Institute for Advanced Study at Harvard University, Cambridge, MA 02138, USA}

\author{I. Baliukin \altaffilmark{3}$^{,}$\altaffilmark{4}$^{,}$\altaffilmark{5}}
\affil{\altaffilmark{3}Space Research Institute of Russian Academy of Sciences, Profsoyuznaya Str. 84/32, Moscow, 117997, Russia}
\affil{\altaffilmark{4}Moscow Center for Fundamental and Applied Mathematics, Lomonosov Moscow State University, GSP-1, Leninskie Gory, Moscow, 119991, Russia}
\affil{\altaffilmark{5}HSE University, Moscow, Russia}

\author{M. Gkioulidou \altaffilmark{6}}
\affil{\altaffilmark{6}Applied Physics Laboratory, Johns Hopkins University, 
Laurel, MD 20723, USA}

\author{J. D. Richardson \altaffilmark{7}}
\affil{\altaffilmark{7}Kavli Institute for Astrophysics and Space Research and Department of Physics, Massachusetts Institute of Technology, Cambridge, MA, USA}

\author{G. P. Zank \altaffilmark{8}}
\affil{\altaffilmark{8}Department of Space Science, The University of Alabama in Huntsville
4 Huntsville, AL 35805, USA}

\author{A. T. Michael\altaffilmark{6}}

\author{G. T{\'o}th\altaffilmark{9}}
\affil{\altaffilmark{9}University of Michigan, Ann Arbor, MI 48109, USA}

\author{V. Tenishev\altaffilmark{9}}

\author{V. Izmodenov\altaffilmark{3}$^{,}$\altaffilmark{4}$^{,}$\altaffilmark{10}}
\affil{\altaffilmark{10}Institute for Problems in Mechanics, Vernadskogo 101-1, Moscow, 119526, Russia}

\author{D. Alexashov\altaffilmark{10}}

\author{S. Fuselier\altaffilmark{11}}
\affil{\altaffilmark{11}Southwest Research Institute, San Antonio, TX, 78228, USA}

\author{J. F. Drake\altaffilmark{12}}
\affil{\altaffilmark{12}Department of Physics and the Institute for Physical 
Science and Technology, University of Maryland, College Park, MD, USA}

\author{K. Dialynas \altaffilmark{13}}
\affil{\altaffilmark{13}Office of Space Research and Technology, Academy of Athens, 10679 Athens, Greece}

\begin{abstract}
Global models of the heliosphere are critical tools used in the interpretation of heliospheric observations. There are several three-dimensional magnetohydrodynamic (MHD) heliospheric models that rely on different strategies and assumptions. Until now only one paper has compared global heliosphere models, but without magnetic field effects. We compare the results of two different MHD models, the BU and Moscow models. Both models use identical boundary conditions to compare how different numerical approaches and physical assumptions contribute to the heliospheric solution. Based on the different numerical treatments of discontinuities, the BU model allows for the presence of magnetic reconnection, while the Moscow model does not. Both models predict collimation of the solar outflow in the heliosheath by the solar magnetic field and produce a split-tail where the solar magnetic field confines the charged solar particles into distinct north and south columns that become lobes. In the BU model, the ISM flows between the two lobes at large distances due to MHD instabilities and reconnection. Reconnection in the BU model at the port flank affects the draping of the interstellar magnetic field in the immediate vicinity of the heliopause. Different draping in the models cause different ISM pressures, yielding different heliosheath thicknesses and boundary locations, with the largest effects at high latitudes. The BU model heliosheath is 15\% thinner and the heliopause is 7\% more inwards at the north pole relative to the Moscow model. These differences in the two plasma solutions may manifest themselves in energetic neutral atom measurements of the heliosphere.
\end{abstract}


\section{Introduction}
The heliosphere is formed through the interaction between the solar wind and the interstellar medium (ISM) as the Sun moves through its local environment. Traditionally, the heliosphere has been regarded as having a long “comet-like” tail extending for thousands of AU \citep{Parker61,Baranov71,Baranov93, Pauls95, Zank96}. This view is being challenged by modeling and observations. \citet{Opher15} suggest that the heliosphere has a “croissant-like” shape instead of a “comet-like” tail, with a short heliotail at low latitudes that only extends 300-400 AU beyond the termination shock. A key feature of this model is that the tension of the solar magnetic field resists stretching by the solar wind and confines the slow solar wind in the inner heliosheath into two jets. Several papers \citep{Izmodenov15,Izmodenov18,Pogorelov15} see evidence for this confinement, but argue for a heliosphere with a long tail that extends for thousands of AU beyond the termination shock.  One key difference between the models of \citet{Izmodenov15} and \citet{Opher15} is that the latter includes non-ideal effects, such as magnetic reconnection (which the numerical approach allows to develop). There are also MHD instabilities along the axis of the solar magnetic field in the heliosheath in \citet{Opher21} which are absent in the works of \citet{Izmodenov15} and \citet{Izmodenov20}. Here we address how these differences give rise to changes in the heliospheric structure.

With a number of different three-dimensional (3D) magnetohydrodynamic (MHD) models being used in the community to simulate and understand the heliosphere, it is critical to benchmark the differences between the models and to understand how the model assumptions affect the results. Previously, \citet{Muller08} compared four hydrodynamic models to explore differences and commonalities arising from different modeling strategies. They found model predictions to be similar in the nose direction, with predicted termination shock distances all within 7\% in the nose and within 14\% in the downwind direction. Additionally, the nose heliopause location agreed within 5\% for each of the models. The main discrepancies arose with regard to the filtration of neutral hydrogen (H) from the ISM and the strength of the interstellar bow shock. 

Although \citet{Muller08} compared the boundary locations for different hydrodynamic models, a comparison of the heliospheric boundaries using models that include the solar and interstellar magnetic fields has not been conducted until now. \citet{Opher17} found that reconnection between the interstellar and solar magnetic fields can lead to a twisting of the interstellar magnetic field around the heliopause. In the model of \citet{Izmodenov15} and \citet{Izmodenov20} (herein referred to as the Moscow model), reconnection between the interstellar and solar magnetic fields at the heliopause is not allowed by their numerical approach. 

In this work, we compare the recently developed SHIELD model \citep{Michael21} (herein referred to as the BU model), which extends the MHD model from \citet{Opher15} to include a kinetic treatment of neutrals, with the MHD model from \citet{Izmodenov15} and \citet{Izmodenov20}. There are other 3D MHD models of the heliosphere \citep{Florinski08,Washimi11,Zank13,Pogorelov15, Heerikhuisen19,Guo19}, but here we focus on the comparison of the BU and Moscow models to understand the effects of non-ideal MHD processes on the heliosphere. These effects are not explicitly suppressed in BU model but are suppressed in the Moscow model. We therefore investigate the difference which arise from the BU model including non-ideal MHD processes while the Moscow model does not. For both models, we use the same boundary conditions. In Section \ref{sec:Model}, we describe the two models and the boundary conditions. In Section \ref{sec:results}, we compare the global structure of the heliosphere and the heliosheath properties for the two models. In Section \ref{sec:Summary}, we present a summary of our results and conclusions.

\section{Model} \label{sec:Model}
\begin{figure*}[t!]
\centering
  \includegraphics[scale=0.5]{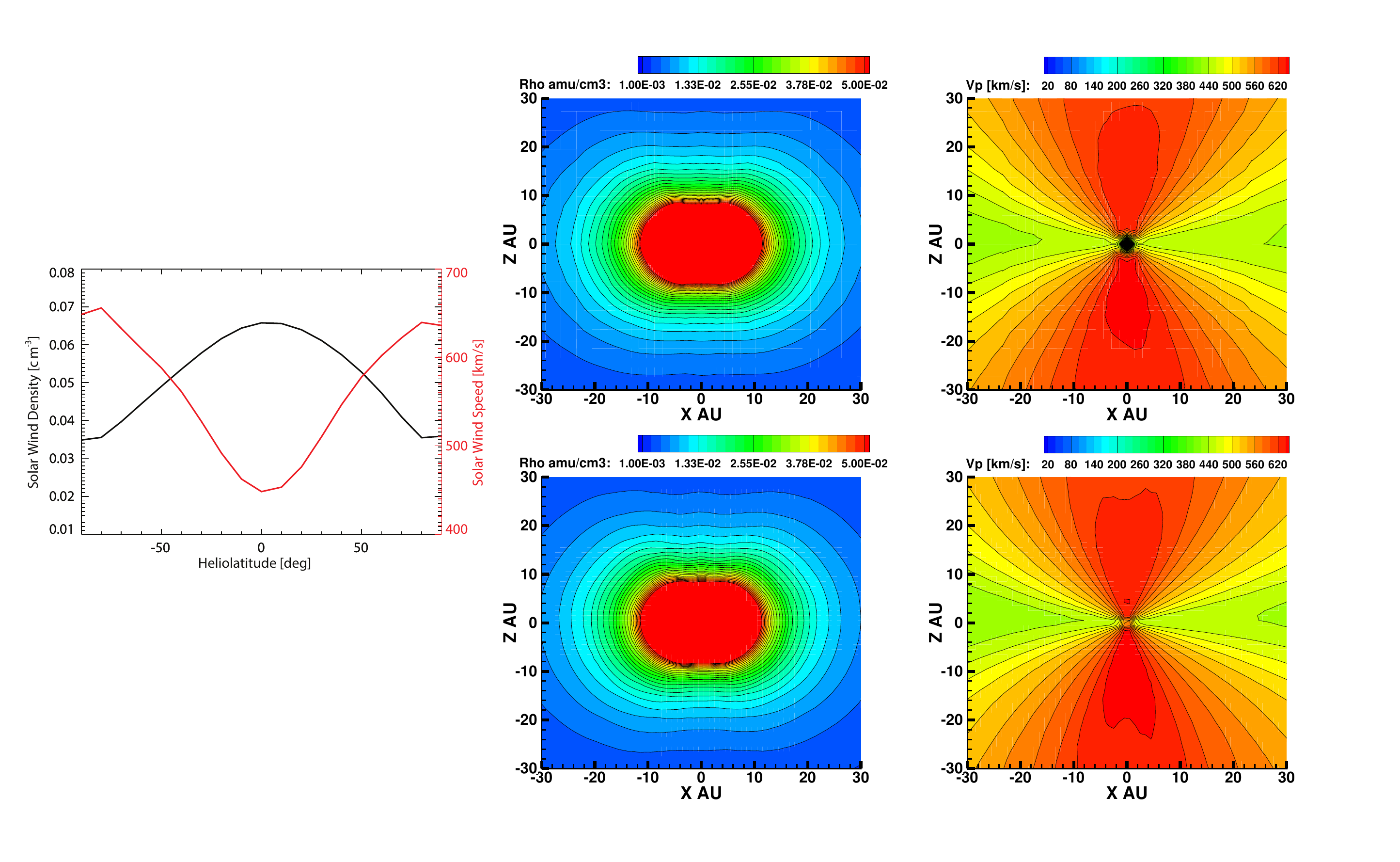}
  \caption{Left column: Boundary conditions at 10 AU for the BU and Moscow models of the solar wind density (black) and speed (red) as a function of heliolatitude. The inner boundary conditions reflect 22-year averaged solar cycle data from 1995-2017. Middle column: Solar wind plasma density out to 30 AU in the meridional plane for the BU (top) and Moscow (bottom) models. Right column: Solar wind plasma speed out to 30 AU in the meridional plane for the BU (top) and Moscow (bottom) models.}
  \label{fig:bcs}
\end{figure*}

The two numerical models of the heliosphere used in this work are the Moscow model \citep{Izmodenov15,Izmodenov20} and the BU model \citep{Opher15,Michael21}. Both models are 3D kinetic-MHD models which treat the cold solar wind plasma and hot pick-up ion plasma as a single fluid, while treating neutral particles kinetically. Both models use the same inner boundary conditions for the solar wind (Figure \ref{fig:bcs}), an averaged solar cycle condition from the years 1995 to 2017 and the same outer boundary conditions for the interstellar neutral H and protons \citep{Izmodenov20}. 

\subsection{Moscow Model} \label{ssec:Moscow}
The Moscow model treats the partially ionized interstellar plasma as a two-component gas consisting of neutral H atoms and a charged plasma consisting of protons, electrons, and helium ions (He+). In the solar wind, the plasma consists of protons, electrons, and alpha particles (He++). 

The neutral atoms are treated kinetically, while the plasma is described via the ideal MHD equations. Source terms for the MHD equations are calculated via the kinetic treatment of neutrals, as integrals of the H-atom velocity distribution are calculated using a Monte Carlo method \citep{Malama91} which solves the kinetic equation (e.g. \citet{Izmodenov01}). The velocity distribution of the proton component is assumed to be locally Maxwellian, though the method allows for a generalization to any isotropic distribution function (see \citet{Malama06}). 

A global iteration method \citep{Baranov93} gives a self-consistent steady-state solution. A 3D moving grid is used to fit discontinuities. A fitting technique proposed by \citet{Godunov79} allows for a fitting of all the major discontinuities present in the simulation – the heliopause, termination shock and bow shock (when applicable). A specific, non-regular moving grid allows for an exact fitting of the termination shock and the heliopause by increasing the resolution and the number of cells in the vicinity of discontinuities (see Figure 2 of \citet{Izmodenov15}). In the non-regular grid, one surface of the grid is chosen to be the heliopause, i.e. tangential discontinuity. The mathematical conditions at the surface are enforced to be exactly: 1) no mass flux through the surface, 2) $B_{n}$=0, and 3) balance of total pressure on the two sides of the surface. With the termination shock and heliopause existing as grid surfaces, the Moscow model is able to define different populations of neutral H atoms based on their place of origin (see, e.g. \citet{Malama91}, \citet{Baranov98}). Optimization of the critical weights for the Monte Carlo method are performed for each of the four neutral H populations separately \citep{Malama90}, allowing for an increase in the efficiency of the method. Additionally, the nature of the grid at the heliopause allows no direct communication (such as magnetic reconnection) between the solar wind and interstellar medium, which is a key difference between the Moscow and BU models. In this work, we interpolate the results from the Moscow model onto a Cartesian grid with uniform resolution of 1 AU for our analysis. The cells of the original grid in the Moscow model generally have 8 vertices, in which the gas-dynamic values are set. Each cell is divided into 6 tetrahedrons, which densely fill the space. To determine the values in an arbitrary point of space, linear interpolation is performed in the corresponding tetrahedron. Using this procedure, the values for the 1 AU Cartesian grid were obtained. The original grid for the Moscow model is shown in Figure 2 of \citet{Izmodenov15}.

The Moscow model solves the ideal MHD equations using a unipolar magnetic field configuration. As noted in \citet{Izmodenov15}, the sign of the radial component of the solar magnetic field is not needed because the terms in the ideal MHD equations which are responsible for the influence of the magnetic field do not depend on its orientation. The solutions do not depend on the orientations of the solar and interstellar magnetic fields along their magnetic field lines. Once the solution is obtained, the polarity of the solar magnetic field can be calculated based on the work of \citet{Barsky99}.

The Moscow model uses the system of coordinates connected with the interstellar flow and magnetic field vectors, as described in \citet{Izmodenov15}. The z-axis is directed toward the interstellar flow. The x-axis is in the plane containing interstellar flow and magnetic field vectors (BV—plane) and perpendicular to the z-axis. The direction of the x-axis is chosen such that the projection of interstellar magnetic field to the x-axis is negative. The y-axis completes the right-handed system of the coordinates. 

\subsection{BU Model} \label{ssec:BU}
The BU model, also known as the Solar-wind with Hydrogen Ion Exchange and Large-scale Dynamics (SHIELD) model, is a numerical model based on the work of \citet{Michael21}. This model extends the MHD model of \citet{Opher15} to include a kinetic treatment of neutral H atoms instead of a multi-fluid treatment of H atoms \citep{Zank96}. The BU model uses the Space Weather Modeling Framework (SWMF) \citep{Toth05} and couples the Outer Heliosphere (OH) and Particle Tracker (PT) components. The OH component is based on the Block-Adaptive Tree Solar wind Roe-Type Upwind Scheme (BATS-R-US) solver \citep{Toth12}, which is a three-dimensional (3D), block adaptive, upwind finite-volume MHD code that is highly parallel. \citet{Opher03} adapted BATS-R-US to the outer heliosphere as the OH component, which is a global 3D multi-fluid simulation. As a standalone component within SWMF, the OH component is capable of treating multiple ion fluids in addition to multiple neutral fluids \citep{Opher20}, but here we use only a single-ion fluid. In a multi-fluid treatment of the heliosphere, the ideal MHD equations are solved for the ion species and Euler’s equations are solved separately for the individual neutral H populations, with each neutral population corresponding to a different region of the heliosphere \citep{Zank96}. Source terms from \citet{McNutt98} connect the ion and neutral fluids, where we treat the thermal solar wind ion and pick-up ions (PUIs) as a single fluid population. 

The PT component of SWMF is based on the Adaptive Mesh Particle Simulation (AMPS) to treat neutral atoms kinetically. AMPS is a global, kinetic, 3D kinetic particle code which solves the Boltzmann equation using a Direct Simulation Monte Carlo method \citep{Tenishev21}. The BU model couples the PT component to the OH component, and in so doing the PT component is used to solve the Boltzmann equation for neutral H atoms streaming through the domain and only incorporates effects due to charge exchange. Based on the work of \citet{Malama91}, a statistical estimation of the momentum and energy source terms is found by summing the change in these quantities from the individual charge exchange events in a given cell over a specific time interval \citep{Michael21}. The neutral H atoms are the only modeled neutral population, and are injected at the outer boundary with a Maxwell-Boltzmann distribution.

When running the BU model, the OH standalone component is used first with the multi-fluid neutral approximation. This is done to relax the plasma to a steady state solution in the MHD model \citep{Opher09,Opher15}. Once steady state is achieved, the OH component is coupled with the PT component and neutral atoms are modeled kinetically. The BU model cycles between the OH and PT components as source terms are passed from the PT component to the OH component. Approximately 148 million particles are modeled within AMPS, and the source terms from charge exchange for the MHD equations accrue for 5000 time steps in the PT component in order to obtain sufficient statistics before being passed to the MHD solver to update the plasma solution in the OH component. When a charge exchange event occurs in the PT component, the MHD parameters from the OH component are interpolated to the location of the particle in the event via linear interpolation. Source terms are accumulated between couplings with the OH component.

The computational domain for both the OH and PT components is x={$\pm$}1500 AU, y={$\pm$}2000 AU, and z={$\pm$}2000. The coordinate system is such that the z-axis is parallel to the solar rotation axis and the x-axis is 5$^{\circ}$ above the direction of interstellar flow, with y completing the right-handed coordinate system. For AMPS in the PT component, the grid resolution within the heliosphere is 4.7 AU from x=-280 AU to 560 AU, y=-500 AU to 500 AU, and z=-380 AU to 380 AU. Beyond this region, the grid resolution gradually decreases until it reaches 18.8 AU at the domain edges in the ISM. For BATS-R-US in the OH component, the resolution is similar to that of \citet{Opher15} and \citet{Michael21}. We use a resolution of 1 AU (from x=-190 to -110 AU, y=-100 to 100 AU, z=-85 to 85 AU) and 2 AU (from x=-190 to -110 AU, y=-300 to 300 AU, z=-200 to 200 AU) at the nose of the heliosphere, and 4 AU resolution (from x=-300 to 1000 AU, y=-410 to 410 AU, z=-450 to 450 AU) in the heliosheath. Both the OH and PT components interpolate the results from the other components when coupled using a bilinear interpolation technique and the interpolated results are stored in cell centers. 

As in the Moscow model, the BU model uses a unipolar solar magnetic field configuration in both hemispheres, as in \citet{Opher15}. The unipolar treatment of the solar magnetic field eliminates spurious numerical effects due to numerical diffusion and reconnection of the solar magnetic field across the heliospheric current sheet \citep{Michael18}. The BU model uses a multi-fluid description to generate the initial state, which requires defining a criteria to identify the different regions where neutrals are created and destroyed separated by boundaries like the heliopause and termination shock so that one knows how to model the multiple neutral fluids. After the final multi-fluid state is obtained, the kinetic description is used which is independent of any criteria and the boundaries form naturally. Several criteria, such as the iso-temperature criteria for the heliopause, are used only in post-processing to try to best capture the iso-surface of the heliopause. It is true that this iso-surface best identifies the forward hemisphere of the heliosphere and is less effective in the tail region. While preventing magnetic dissipation within the heliosheath, the unipolar magnetic field allows reconnection to occur in the port side (defined as the left side of the heliosphere as it moves through the ISM as seen within looking out) and tail of the BU model. \citet{Opher17,Opher20} suggest that this reconnection explains the draped magnetic field ahead of the heliopause as revealed by Voyager 1 and 2 data. Using a dipolar solar magnetic field, \citet{Michael18} showed that reconnection across the current sheet with the interstellar magnetic field at the heliopause leads to a draped interstellar magnetic field that instead deviates from observations. In the BU model the solar and interstellar magnetic field lines are able to reconnect, but in the Moscow model no communication is allowed at the heliopause. 

\subsection{Boundary Conditions} \label{ssec:BCs}
To compare the BU and Moscow models, we use identical inner and outer boundary conditions taken from Model 1 of \citet{Izmodenov20}. A difference between the two models is that the BU model does not include He ions or alpha particles, but the Moscow model does. For this study we ran the Moscow model with no He ions or alpha particles
so the model comparison is valid. In the interstellar medium, the proton density is assumed to be $n_{p,ISM}$ = 0.04 cm$^{-3}$, while the neutral H atom density is $n_{H,ISM}$ = 0.14 cm$^{-3}$. All of the neutral and ionized populations in the interstellar medium are assumed to have the same bulk velocity $v_{ISM}$ = 26.4 km/s (longitude = 75.4$^{\circ}$, latitude = -5.2$^{\circ}$ in ecliptic J2000 coordinate system) and temperature $T_{ISM}$ = 6530 K at the outer boundary, where there is pristine ISM not mediated by the heliosphere. We use the interstellar magnetic field intensity and orientation corresponding to $B_{ISM}$ = 3.75 $\mu$G and $\alpha$ = 60$^{\circ}$, where the magnetic field is aligned with the hydrogen deflection plane \citep{Lallement05} and $\alpha$ is the angle between the interstellar velocity and magnetic field vectors. 

For the inner boundary conditions, we use 22-year averaged solar cycle conditions (1995-2017) as in \citet{Izmodenov20}. Helio-latitudinal variations of the solar wind density and speed are taken into account, and the temperature is related to the speed via the Mach number. At Earth, a Mach number of M=6.44 (corresponding to a solar wind temperature of $T_{SW}$ = 188500 K) is used. In the ecliptic plane, hourly-averaged solar wind data from the OMNI 2 dataset is used for the density and speed. Heliolatitudinal variations of the solar wind speed are based on analysis of interplanetary scintillation (IPS) observations \citep{Tokumaru12} from 1990 to 2017. For heliolatitudinal variations of the solar wind mass flux, SOHO/SWAN full-sky maps of backscattered Lyman-alpha intensities are used \citep{Quemerais06, Lallement10,Katushkina13, Katushkina19}. Data from SOHO/SWAN are available from 1995 to the end of 2017, and an inversion procedure is used to obtain solar wind mass flux as a function of time and heliolatitude. For the solar magnetic field, a Parker solution is assumed for both models, with the radial component of the magnetic field as $B_{SW}$ = 37.5 $\mu$G at 1 AU. 

For the Moscow model, the inner boundary conditions are implemented at 1 AU, while in the BU model the inner boundary conditions are implemented at 10 AU. The boundary conditions are matched at 10 AU by extracting the solar wind conditions from the Moscow model at that distance and implementing them in the BU model. At 10 AU in the ecliptic plane, the plasma density is 0.067 cm$^{-3}$ and the plasma speed is 442 km/s. In Figure \ref{fig:bcs}, we present the latitudinal profile of the solar wind proton density and speed at 10 AU, which is assumed to be independent of longitude, and the comparison of the solar wind conditions out to 30 AU generated by the two models.

\section{Results}\label{sec:results}

\subsection{Global Structure of the Heliosphere}\label{ssec:global}
One of the major questions that led to the code comparison documented in the present paper is whether the solar magnetic field of the heliosphere is strong enough to collimate the plasma flow into distinct lobes in the downtail region. Both models do collimate the solar wind outflow into distinct lobes. On the other hand, there are distinct differences in spite of both models using the same boundary conditions. The low latitude region of the tail in the BU model is dominated by ISM plasma while the heliopause in the Moscow model prevents penetration of ISM into this region. The reasons for such differences are discussed below.

\begin{figure*}[t!]
\centering
  \includegraphics[scale=0.45]{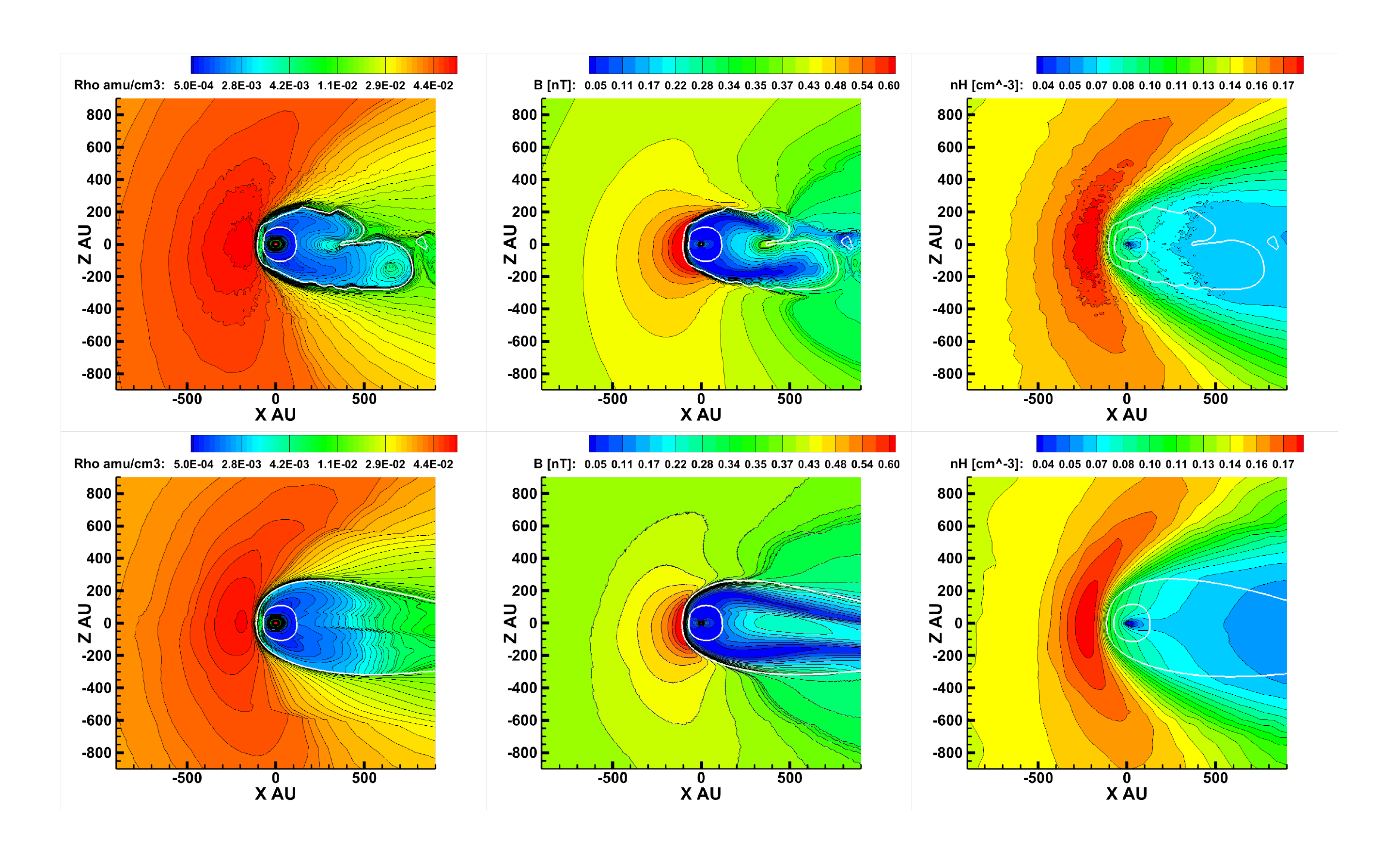}
  \caption{Comparison of BU (top) and Moscow (bottom) solutions of the heliosphere in the meridional plane Included are color and line contours of proton density [cm$^{-3}$] (left), magnetic field intensity [nT] (middle) and neutral H density [cm$^{-3}$] (right).The white lines represent the termination shock (inner) and heliopause (outer) for each model.}
  \label{fig:domain}
\end{figure*}

\begin{figure*}[t!]
\centering
  \includegraphics[scale=0.7]{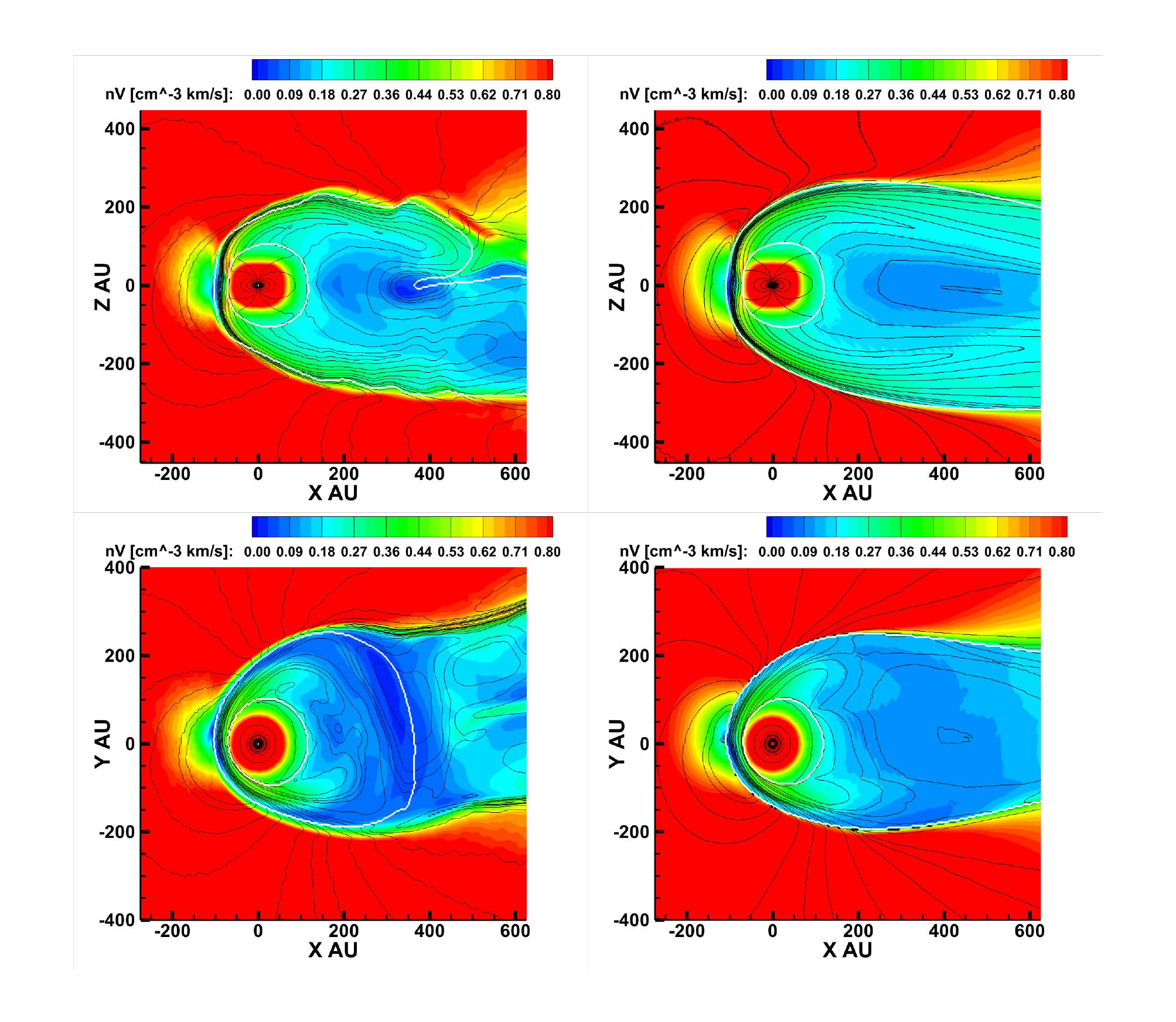}
  \caption{Meridional (top) and equatorial (bottom) slices of BU (left) and Moscow (right) models showing the mass flux ($n_{p}V_{p}$) through the heliosheath. Black lines represent the magnetic field intensity overlaid on the mass flux.}
  \label{fig:massflux}
\end{figure*}

Both models show collimation of the the heliosheath plasma by the solar magnetic field (as shown by \citet{Opher15} and \citet{Drake15}, see also collimation in astrospheres by \citet{Golikov17a,Golikov17b} and \citet{Korolkov21}). Figure \ref{fig:domain} shows a comparison of the two solutions in the meridional plane (defined as the Y=0 AU plane in the BU model coordinate system). Both models produce a split tail in the sense that the solar magnetic field confines the charged solar particles into distinct north and south columns that become lobes (Figure \ref{fig:massflux}). In the BU model, the ISM flows between the two lobes at distances larger than 300 AU from the Sun. There are indications that instabilities that can develop along the axis of the solar magnetic field in the BU model \citep{Opher21}, which are not seen in the Moscow model, are the key reason for the ISM flowing between the two lobes. In the current paper, the BU model is in a quasi-steady state, determined by evaluating whether the location of the termination shock and heliopause in the nose and tail no longer change with time. The quasi-steady state model captures the presence of time dependent features, such as instabilities, which are absent in the Moscow model. The role of instabilities and the time dependence of the solution will be studied in future work when the BU model is extended to be fully time dependent.

The plasma density downstream of the termination shock both in the upwind (nose) direction and downwind (tail) direction within 300 AU of the Sun are similar for both the BU and Moscow models. Beyond 300 AU in the downwind direction, the plasma densities are different in both models since the ISM flows between the two lobes in the BU model (Figure \ref{fig:domain}). In both models, the plasma is confined in two lobes in the tail; however, in the BU model approximately 300 AU from the Sun, there is an enhancement of the plasma density where solar magnetic field lines are open to the ISM. This is due to instabilities \citep{Opher21} present in the BU model.  

The comparison between the magnetic field intensity between the BU and Moscow models is also shown in Figure \ref{fig:domain}. As with the plasma density, the magnetic field in the Moscow model extends down the tail for thousands of AU uninterrupted by the ISM. There is a clear evidence of a split in the heliotail at high and low latitudes by the solar magnetic field in both models. The neutral H solutions are similar for the two models.

In Figure \ref{fig:massflux}, we compare the mass flux from the BU and Moscow models in the meridional plane. Within 100 AU of the termination shock in the heliotail the mass flux between the two models shows a similar profile. Black lines representing the magnetic field intensity are overlaid on the mass flux, which indicate that the mass flux is organized by the solar magnetic field, with higher mass flux at high latitudes and lower mass flux at lower latitudes. Beyond 100 AU from the termination shock, differences in the mass flux of the two models arise in response to the deflection of the solar wind plasma in the BU model due to the shortened heliotail.

\subsection{Interstellar Medium Properties}\label{ssec:ism}

\begin{figure*}[t!]
\centering
  \includegraphics[scale=0.6]{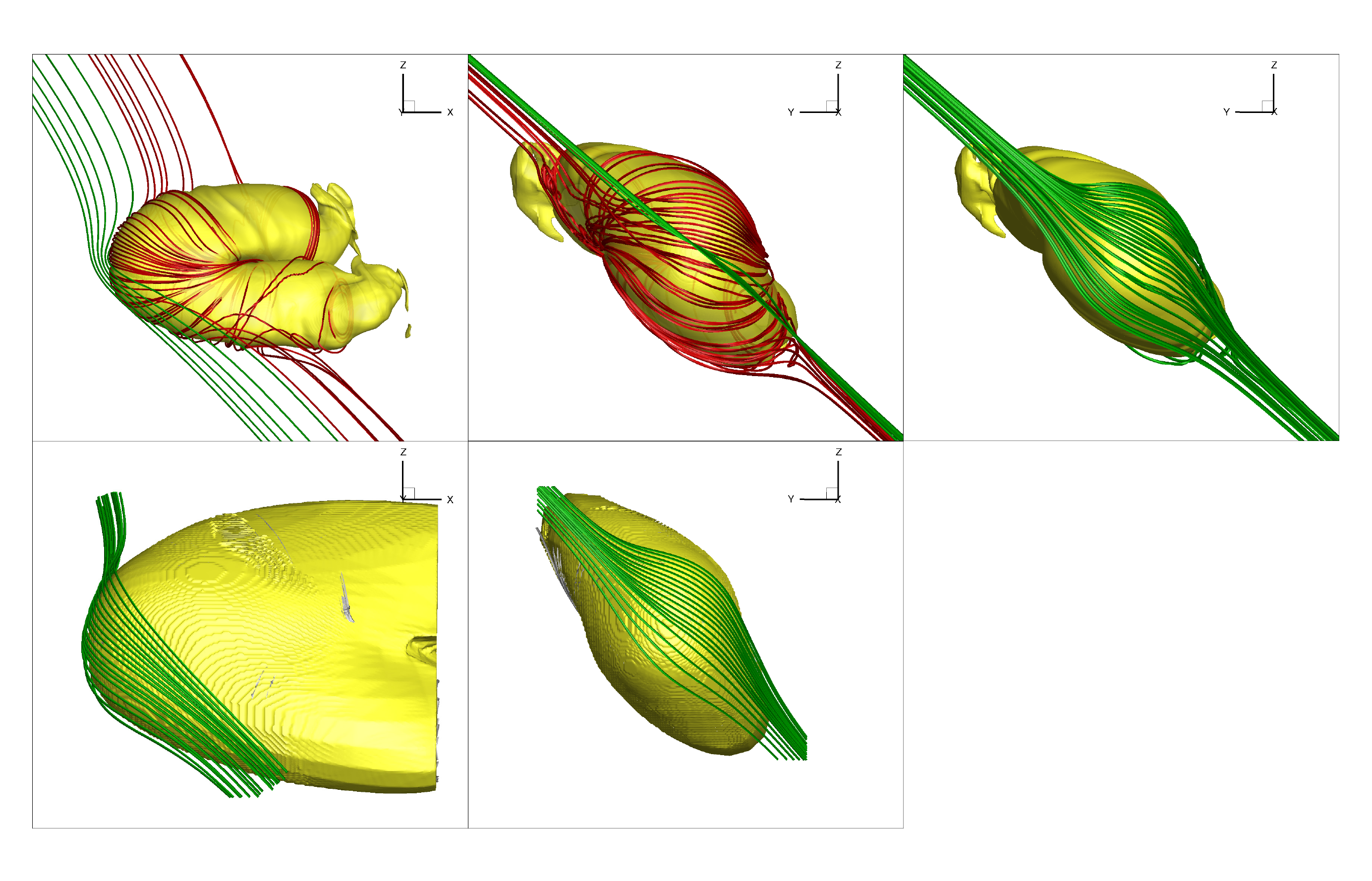}
  \caption{Draping of interstellar magnetic field lines around the heliopause for the BU model (top) and Moscow model (bottom). Left: port-side view of the heliosphere. Middle/Right: view from outside of the heliosphere along the direction of ISM flow. Green lines reflect interstellar magnetic field lines that have not undergone magnetic reconnection. Red lines reflect interstellar magnetic that have undergone magnetic reconnection with solar magnetic field lines. The yellow isosurfaces represent the heliopause, with the heliopause in the BU model represented by the isosurface of lnT = 12.55. The top middle and top right figures show reconnected interstellar magnetic field lines and non-reconnected interstellar magnetic field lines in the BU model draped along the nose of the heliosphere, respectively. Figures are not in the same scale.}
  \label{fig:draping}
\end{figure*}

Upstream of the heliopause in the ISM, the solution looks fairly similar in both models, though there are differences. The differences in the ISM stem from different treatments of the heliopause in the two models. The inclusion of non-ideal MHD effects in the BU model causes the draping of the interstellar magnetic field to be very different between the two models (Fig. \ref{fig:draping}). The red lines reflect reconnected magnetic field lines that originate at the Sun and are open to the ISM, while green lines reflect non-reconnected magnetic field lines. In the Moscow model (bottom panels) the draping of the interstellar magnetic field reflects ideal draping around a surface with the axis of symmetry along the direction of the undisturbed interstellar magnetic field plane. For the BU model, the interstellar magnetic field ahead of the heliopause twists to an azimuthal direction (top panels) close to the heliopause. This twist is due to reconnection in the port flank of the heliosphere \citep{Washimi15}. 

The BU model suppresses reconnection in the nose region while allowing it in the flanks \citep{Opher17}, consistent with recent ideas about reconnection suppression from diamagnetic drifts. Most of the reconnection at the nose occurs between the $B_{y}$ components of the solar and interstellar magnetic fields. By choosing the same polarity of the ISM for the solar magnetic field, the BU model suppresses the reconnection at the nose. The stabilizing effect of diamagnetic drifts is an important kinetic effect missed by MHD models, which can develop at boundaries such as the Earth's magnetospause or the heliopause \citep{Swisdak03,Swisdak10}. The stabilizing influence of these drifts has been extensively documented with solar wind and magnetospheric data \citep{Phan10,Phan13}. The diamagnetic drift suppresses reconnection when the drift speed is larger than the Alfv\'en speed in the reconnecting magnetic field. The jump in plasma $\beta$ (the ratio of the plasma pressure to the magnetic pressure) across the nose of the heliopause is much greater than in the flanks because the heliosheath $\beta$ is greater at the nose than in the flanks \citep{Opher17}. Large-scale reconnection is therefore suppressed in the nose but not at the flanks. The twisting of the interstellar magnetic field to the azimuthal direction ahead of the heliopause leads to an increase in the magnetic field intensity at higher latitudes along the heliopause. In the BU model, there is a build-up of the interstellar magnetic field at the nose of the heliosphere and at high latitudes, as demonstrated in Figure \ref{fig:domain}.

\begin{figure*}[t!]
\centering
  \includegraphics[scale=0.45]{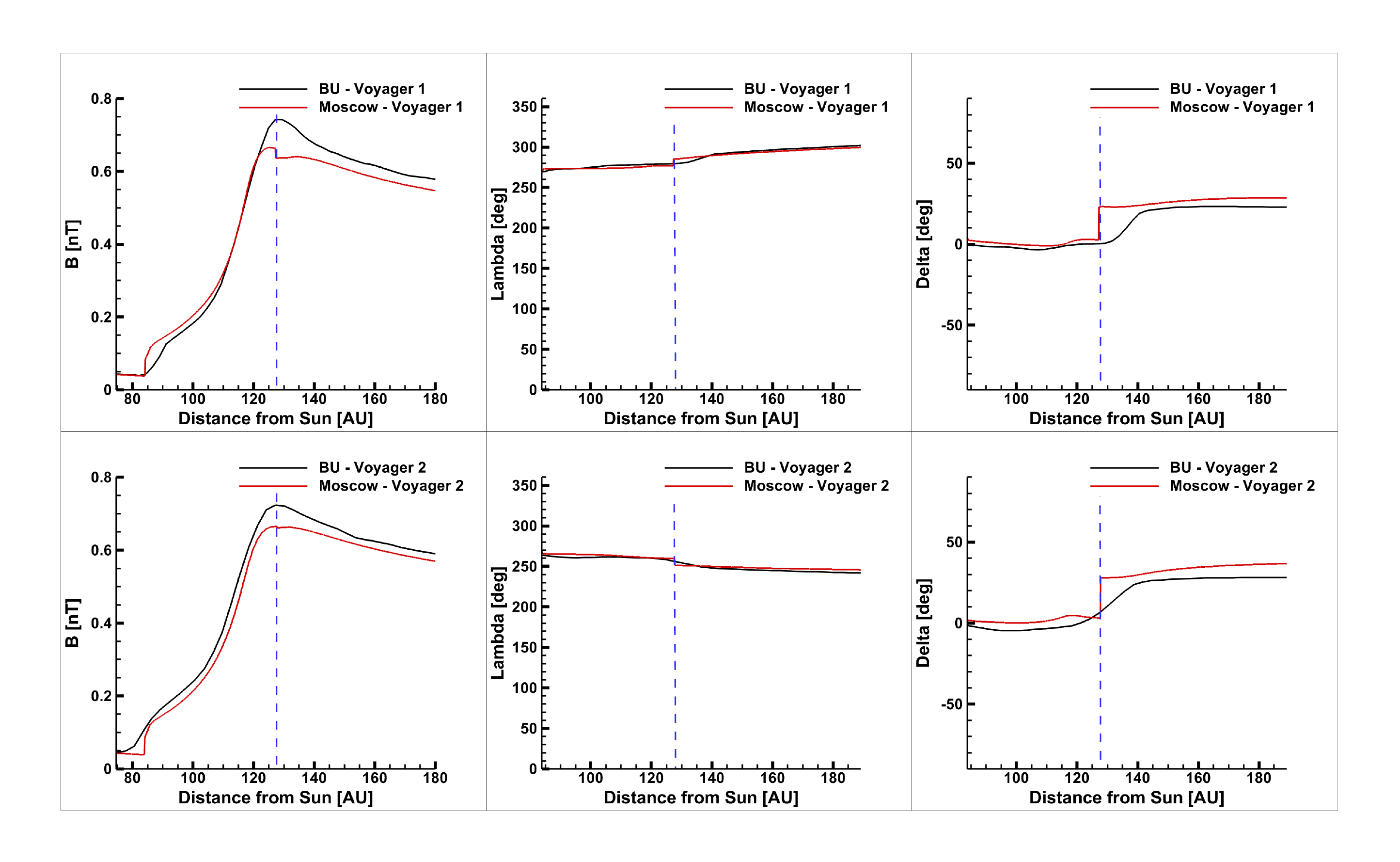}
  \caption{1D cuts along the Voyager 1 (top) and Voyager 2 (bottom) trajectories for the BU (black) and Moscow (red) models. Included are the magnetic field strength (left), azimuthal angle of the magnetic field (middle), and the elevation angle of the magnetic field (right).The elevation angle, $\delta$, is given by ${\delta}=sin^{-1}(B_{N}/B)$ and the azimuthal angle, $\lambda$, is given by ${\lambda}=tan^{-1}(B_{T}/B_{R})$, where the RTN coordinate system is the local Cartesian system centered at the spacecraft. R is radially outward from the Sun, T is in parallel with the plane of the solar equator and is positive in the direction of solar rotation, and N completes a right-handed system. The heliopause is shifted to be at the same location (128 AU) for both models, represented by the dashed vertical blue line.}
  \label{fig:voyagerB180}
\end{figure*}

Figure \ref{fig:voyagerB180} shows the interstellar magnetic field intensity, the elevation angle of the magnetic field ($\delta$), and the azimuthal angle of the magnetic field ($\lambda$) in the Voyager 1 and 2 directions. The elevation angle is given by ${\delta}=sin^{-1}(B_{N}/B)$ and the azimuthal angle is given by ${\lambda}=tan^{-1}(B_{T}/B_{R})$. We use the RTN coordinate system, where R is radially outward from the Sun, T is parallel with the plane of the solar equator and is positive in the direction of solar rotation, and N completes a right-handed system. The termination shock spans approximately 2 grid cells with 4 AU resolution in the BU model, therefore the numerical effects of resolution makes the solar magnetic field intensity appear continuous at the termination shock. At large distances from the heliopause, the profiles of the interstellar magnetic field are very similar. They are not identical due to the sub-alfvenic nature of the flow between the heliopause and the pristine ISM. The differences at the heliopause propagate to larger distances ahead of the heliopause. For the assumed ISM conditions, a fast bow shock is not present in the solutions, likely attributable to neutrals mediating and smoothing the bow shock through charge exchange with the plasma \citep{Izmodenov99,Izmodenov09,Zank13}.

Figure \ref{fig:voyagerB180} shows the magnetic field close to the heliopause for both models. For the BU model, there is a smooth transition in the azimuthal angle ($\lambda$) and the elevation angle ($\delta$) of the magnetic field. As described in \citet{Opher17}, reconnection at the heliopause yields a magnetic field configuration that appears “solar-like” due to the twisting of the magnetic field around the heliopause (similar to the classic Parker spiral). For the Moscow model, where magnetic reconnection is suppressed, there is a discontinuity in both the azimuthal and elevation angles of the magnetic field at the heliopause. The magnetic field change at the heliopause in both models is consistent with Voyager 1 and 2 observations \citep{Burlaga14,Burlaga19}.

\begin{figure*}[t!]
\centering
  \includegraphics[scale=0.45]{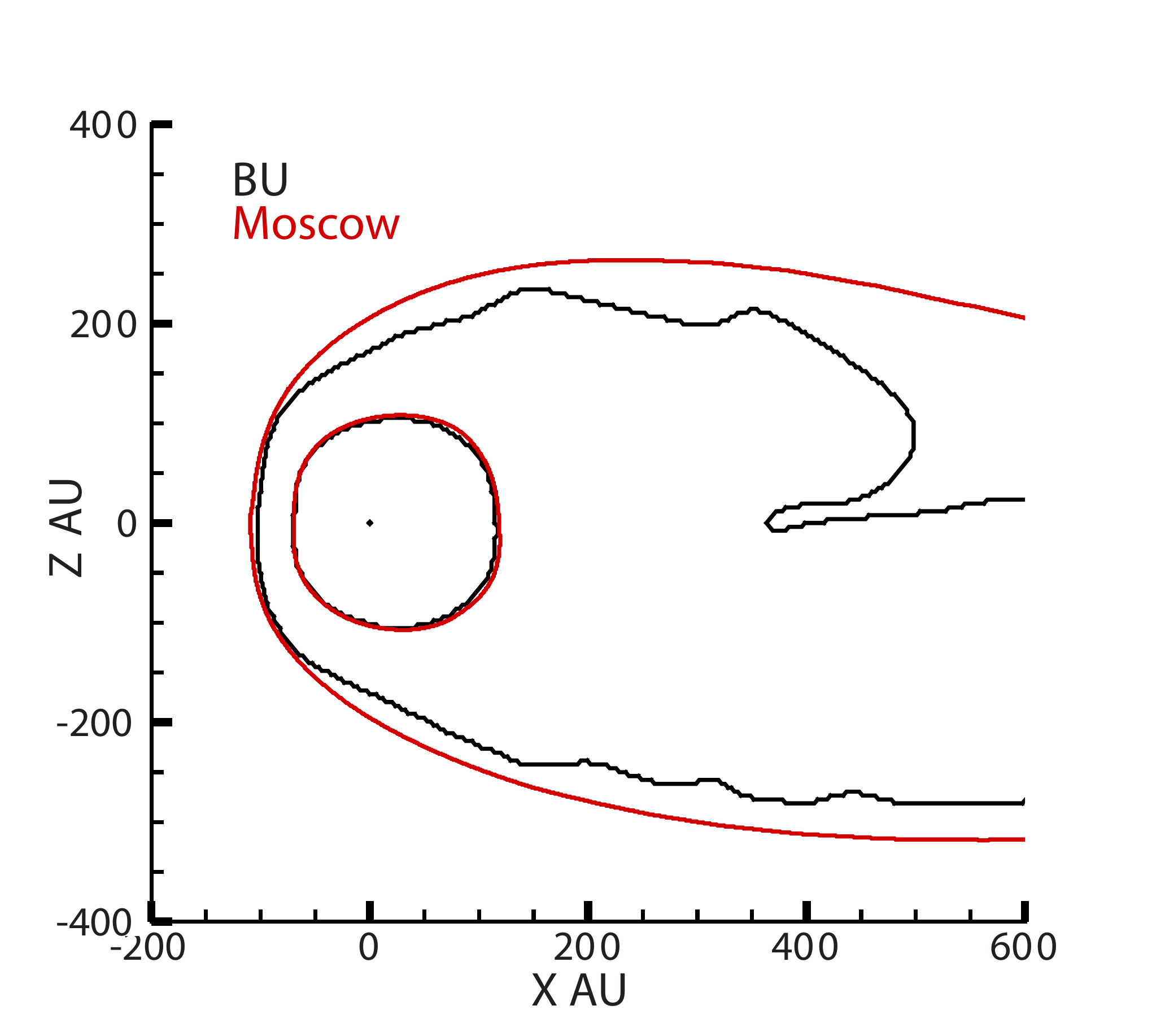}
  \caption{Comparison of heliospheric boundary locations in the BU and Moscow models. Left: comparison of termination shock (inner line) and heliopause (outer line) for the BU (black) and Moscow (red) models. The heliopause for the BU model is defined here in post-processing as lnT = 12.55.}
  \label{fig:boundary}
\end{figure*}

The difference in the pressure at the heliopause affects the distance of the heliospheric boundaries in the two models. As noted in Figure \ref{fig:boundary}, the heliopause is further from the Sun in the Moscow model than in the BU model. The stronger magnetic pressure due to the twisting of the interstellar magnetic field in the BU model leads to a higher overall pressure and therefore more compression of the heliosphere. The twisting of the interstellar magnetic field in the BU model that is absent in the Moscow model primarily affects the heliosphere at high latitudes and in the heliotail. Therefore, the heliopause in the BU model is closer to the Sun than in the Moscow model at high latitudes (7\% at the north pole). When comparing heliopause distances here and throughout this work, we use streamlines to trace the exact location of the boundary in the BU and Moscow models.

\begin{figure*}[t!]
\centering
  \includegraphics[scale=0.45]{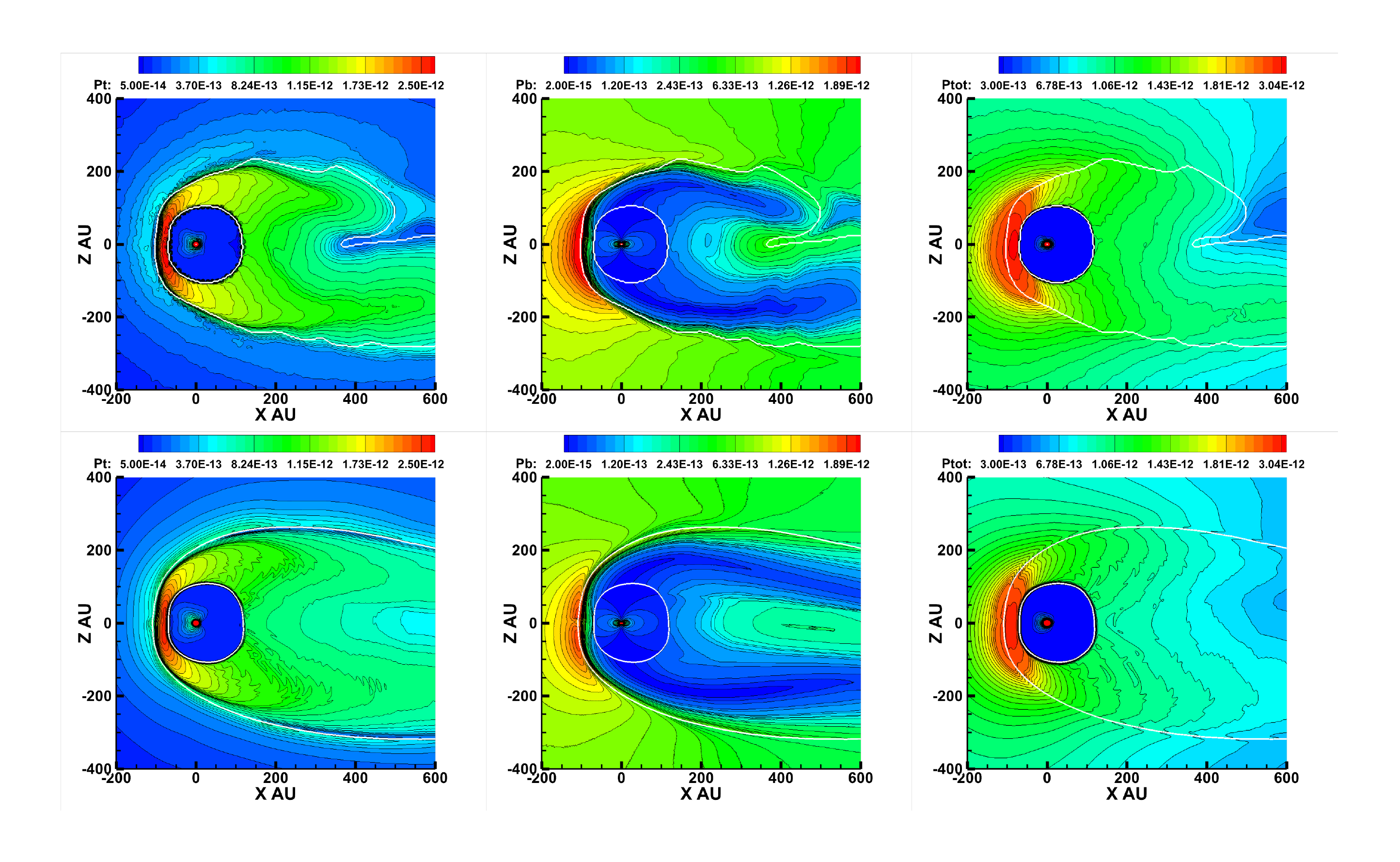}
  \caption{Comparison of pressures in the heliosphere and ISM for the BU (top) and Moscow (bottom) models for the thermal pressure (left), magnetic pressure (middle), and the combined magnetic and thermal pressures (right) in units of dynes/cm$^{2}$. The white lines represent the termination shock (inner) and heliopause (outer) for each model. }
  \label{fig:pressureplane}
\end{figure*}

The magnetic pressure at high latitudes is higher outside the heliopause in the BU model compared to the Moscow model (Figure \ref{fig:pressureplane}). The total pressure outside of the heliopause at high latitudes is also higher in the BU model, giving rise to a more compressed heliosphere. Beyond 350 AU at low latitudes, the ISM flows between the two lobes in the BU model, which leads to the lower thermal pressure and higher magnetic pressure in the tail. Both models have total pressures which agree with Voyager observations \citep{Rankin19, Dialynas20}. 

\begin{figure*}[t!]
\centering
  \includegraphics[scale=0.55]{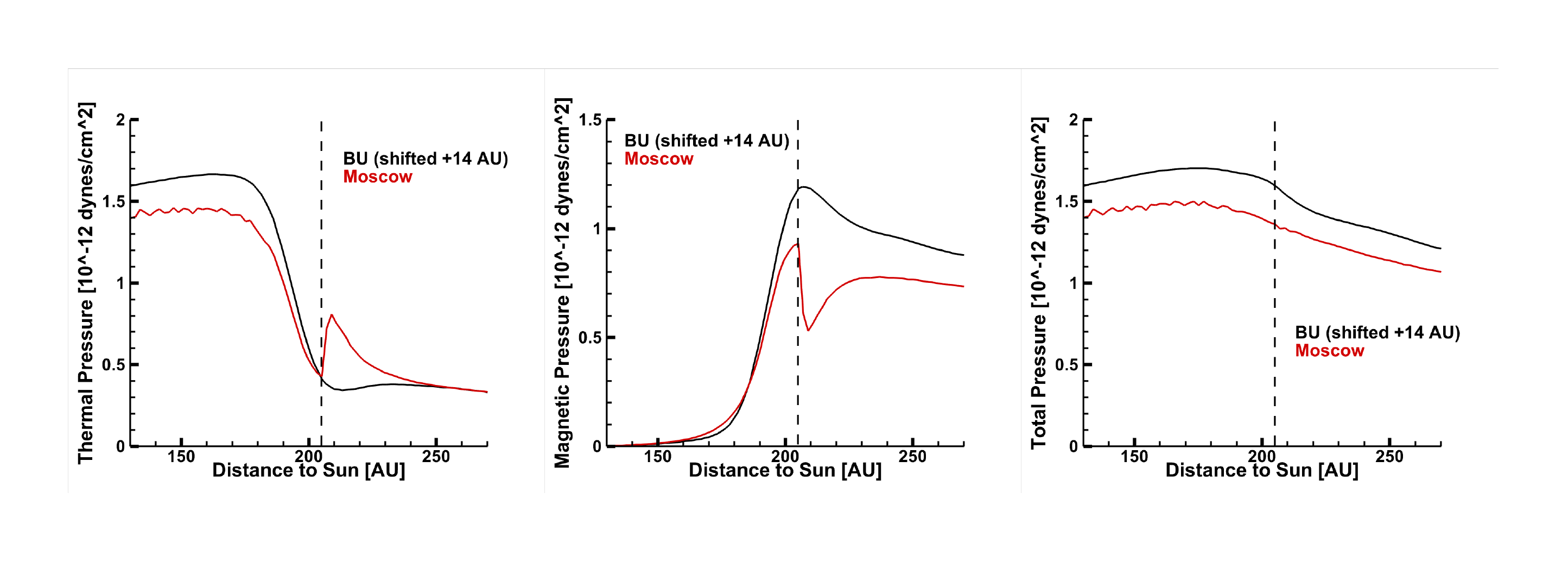}
  \caption{1D cuts of the thermal (left), magnetic (middle), and the combined thermal and magnetic pressures (right) across the heliopause (dashed black vertical line) for the BU (black) and Moscow (red) models at the northern pole (heliolatitude = 90$^{\circ}$). Pressures for the BU model are shifted inward by 14 AU to align the heliopause locations of the two models. }
  \label{fig:pressure1d}
\end{figure*}

Toward the northern pole, the heliosheath is 13 AU thinner (15\%) in the BU model relative to the Moscow model even though the difference in termination shock locations is only 1 AU. Figure \ref{fig:pressure1d} shows 1D cuts near the heliopause of the pressure variation for both models in the northern pole of the heliosphere. We shift the 1D cut for the BU model inwards by 14 AU to align the two heliopause locations. Due to the lack of communication between the ISM and solar wind in the Moscow model, there is a sharp jump in the thermal pressure of the plasma and the magnetic pressure at the heliopause while the magnetic field pressure replaces the plasma pressure so that the total pressure is conserved. This effect is not present in the BU model. For the Moscow model, there is a higher thermal pressure for the plasma just beyond the heliopause as compared to the BU model, but 50 AU past the heliopause the Moscow and BU models have the same thermal pressure. In contrast, the Moscow model has a notably lower magnetic pressure just beyond the heliopause as compared to the BU model, but 50 AU past the heliopause the BU model still maintains a higher magnetic pressure by 17\%. As mentioned previously, this higher magnetic pressure present in the BU model is a consequence of the twisting of the interstellar magnetic field due to reconnection present in the model. This higher magnetic pressure outside of the heliosphere at high latitudes gives rise to a higher total pressure by 13\%, which leads to a compression of the heliosphere relative to the Moscow model.

\subsection{Heliosheath Properties}\label{ssec:heliosheath}
\begin{figure*}[t!]
\centering
  \includegraphics[scale=0.45]{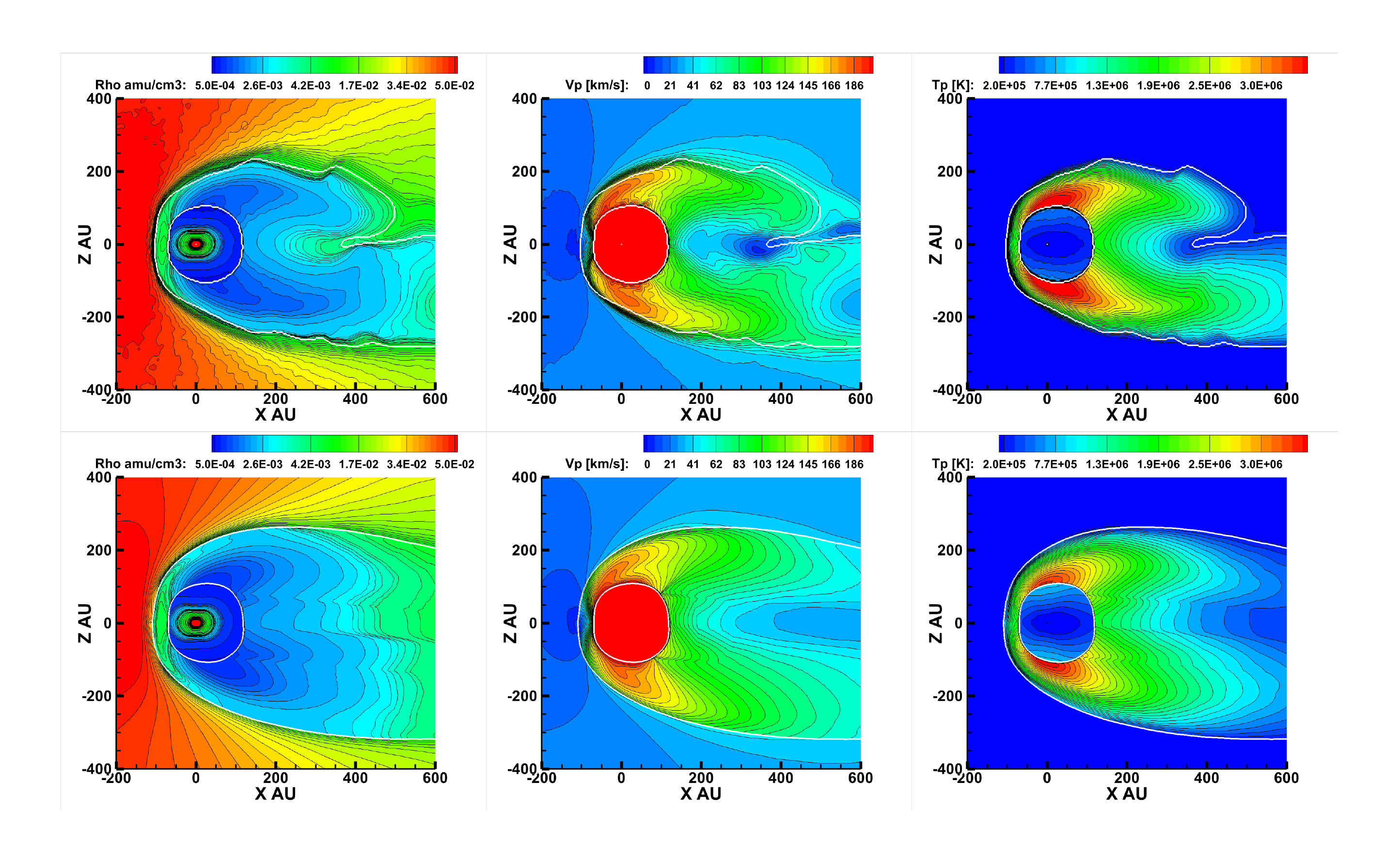}
  \caption{Meridional slices of plasma conditions within the heliosheath for the BU model (top) and the Moscow model (bottom). Presented are color contours and lines of the plasma density [cm$^{-3}$] (left), plasma speed [km/s] (middle), and plasma temperature [K] (right). The white lines represent the termination shock (inner) and heliopause (outer) for each model.}
  \label{fig:heliosheath}
\end{figure*}

\begin{figure*}[t!]
\centering
  \includegraphics[scale=0.45]{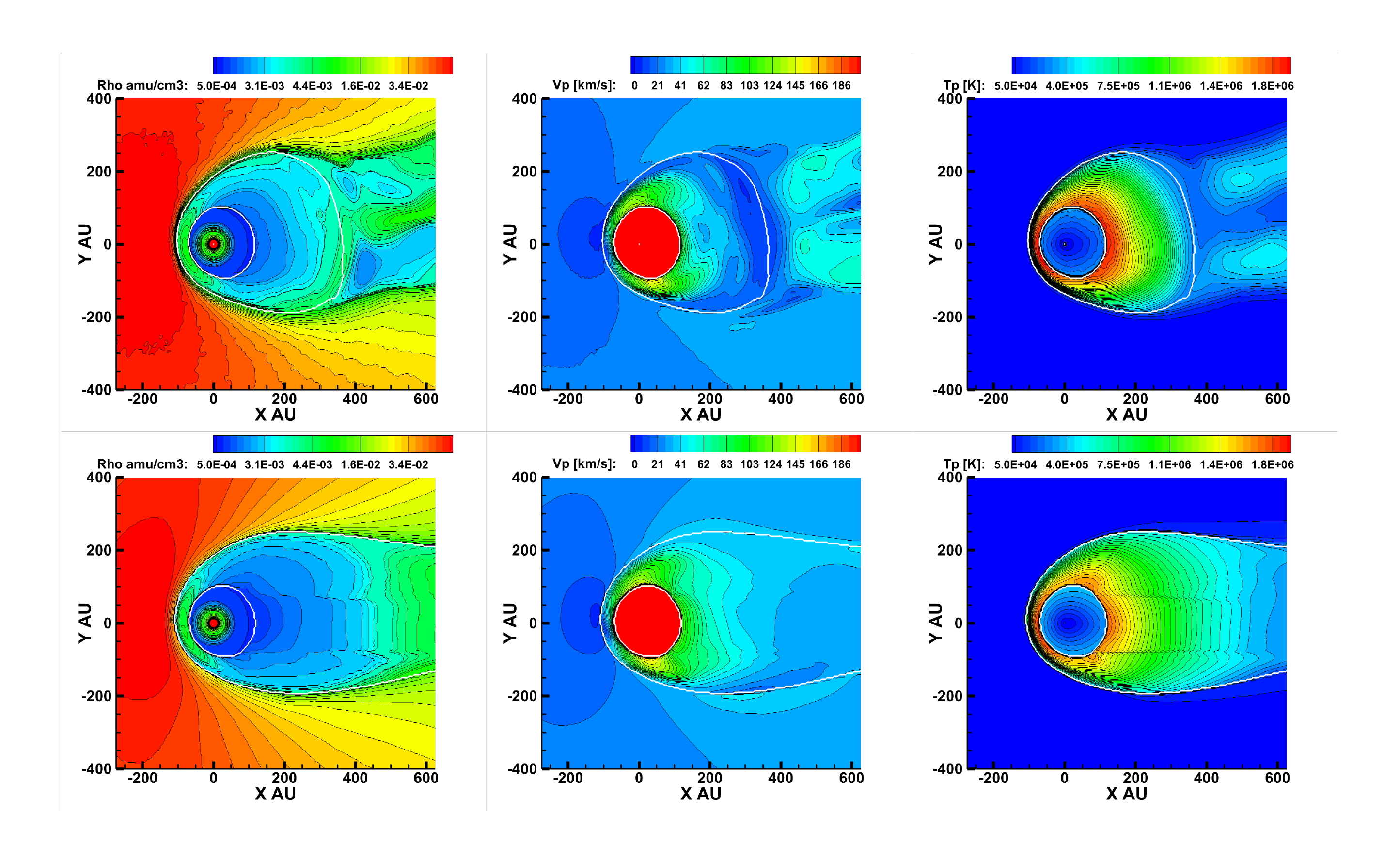}
  \caption{Same as Figure \ref{fig:heliosheath}, but in the equatorial plane.}
  \label{fig:heliosheath_eq}
\end{figure*}

The compression of the heliosphere in the BU model relative to the Moscow model primarily affects the heliopause (Fig. \ref{fig:boundary}). In Figures \ref{fig:heliosheath} and \ref{fig:heliosheath_eq}, we present meridional and equatorial slices, respectively, of the BU and Moscow models showing the plasma density, speed, and temperature in the heliosheath. 

\begin{figure*}[t!]
\centering
  \includegraphics[scale=0.7]{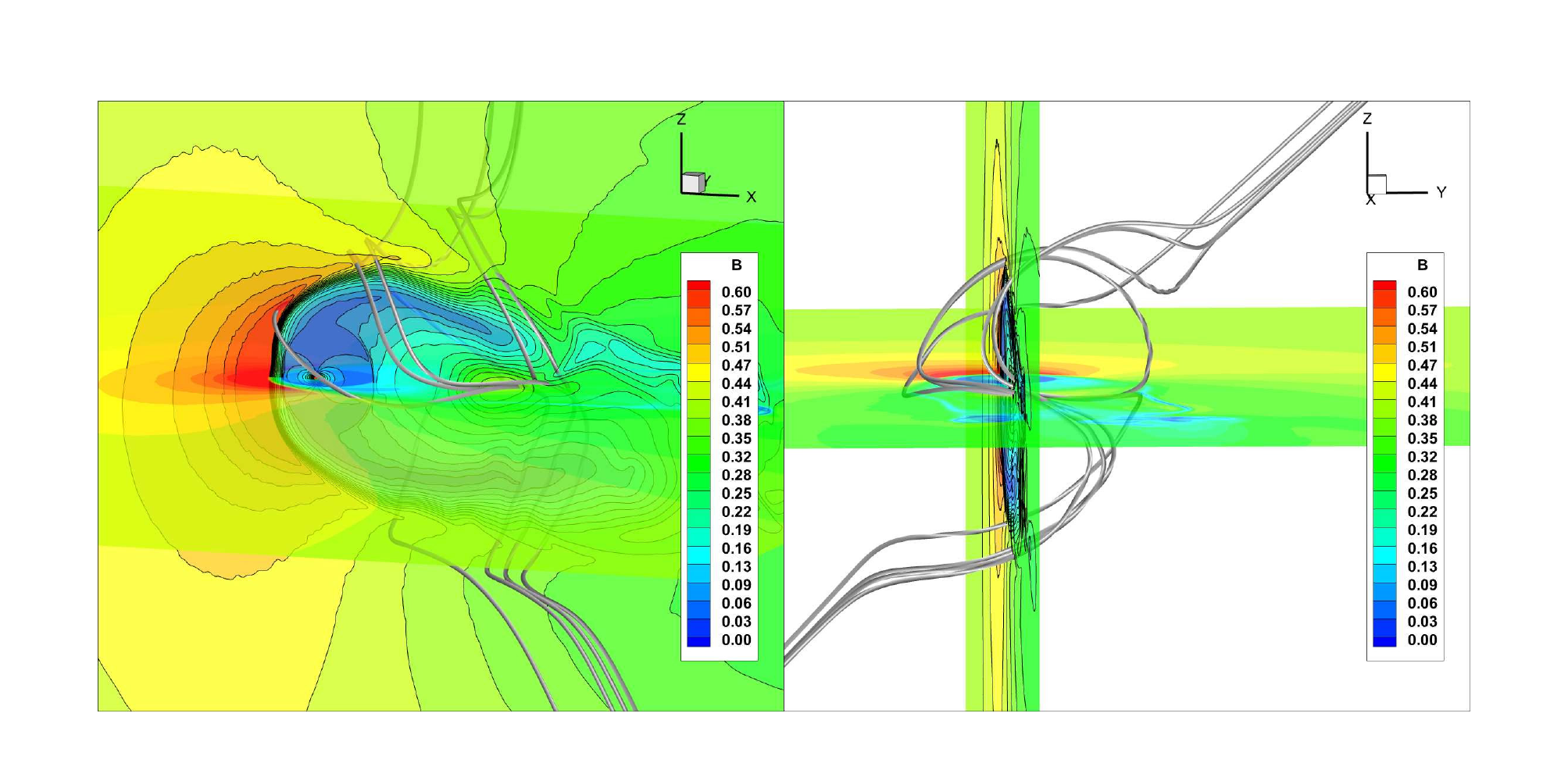}
  \caption{Intersection of the meridional and equatorial planes with contours of magnetic field strength for the BU model. Grey lines reflect reconnected interstellar magnetic field lines passing through a region of mixed ISM and solar wind plasma between the lobes of the BU model. Left: port-view of the reconnected lines through the heliosphere. Right: view from tail towards the nose of the reconnected lines through the heliosphere.}
  \label{fig:Blines}
\end{figure*}

Aside from the split of the heliotail present in the BU model, the properties in the heliosheath for both models are very similar. Within 300 AU of the Sun, the plasma density, speed, and temperature profiles in the heliosheath are similar for both models. The splitting of the tail occurs at roughly 300 AU in the meridional slice. The BU model flows stagnate in this region, which does not occur in the Moscow model. This stagnation region reflects the change of the plasma flow in response to the heliopause down the tail. Beyond 350 AU, where the ISM splits the heliotail, the properties differ between the models. The plasma density and temperature in the BU model reflects a mixing between the heliosheath and ISM plasma in this region. This is also seen in the equatorial speed profiles of the BU model in Figure \ref{fig:heliosheath_eq}, where turbulent mixing is present beyond the stagnation region at 300 AU. This turbulence prevents velocity streamline tracing from the ISM within the region of mixed ISM and solar wind plasma; however, Figure \ref{fig:Blines} shows the intersection of reconnected interstellar magnetic field lines penetrating the region between the lobes of the BU model where the two ends of the field lines are connected to the ISM.

\begin{figure*}[t!]
\centering
  \includegraphics[scale=0.54]{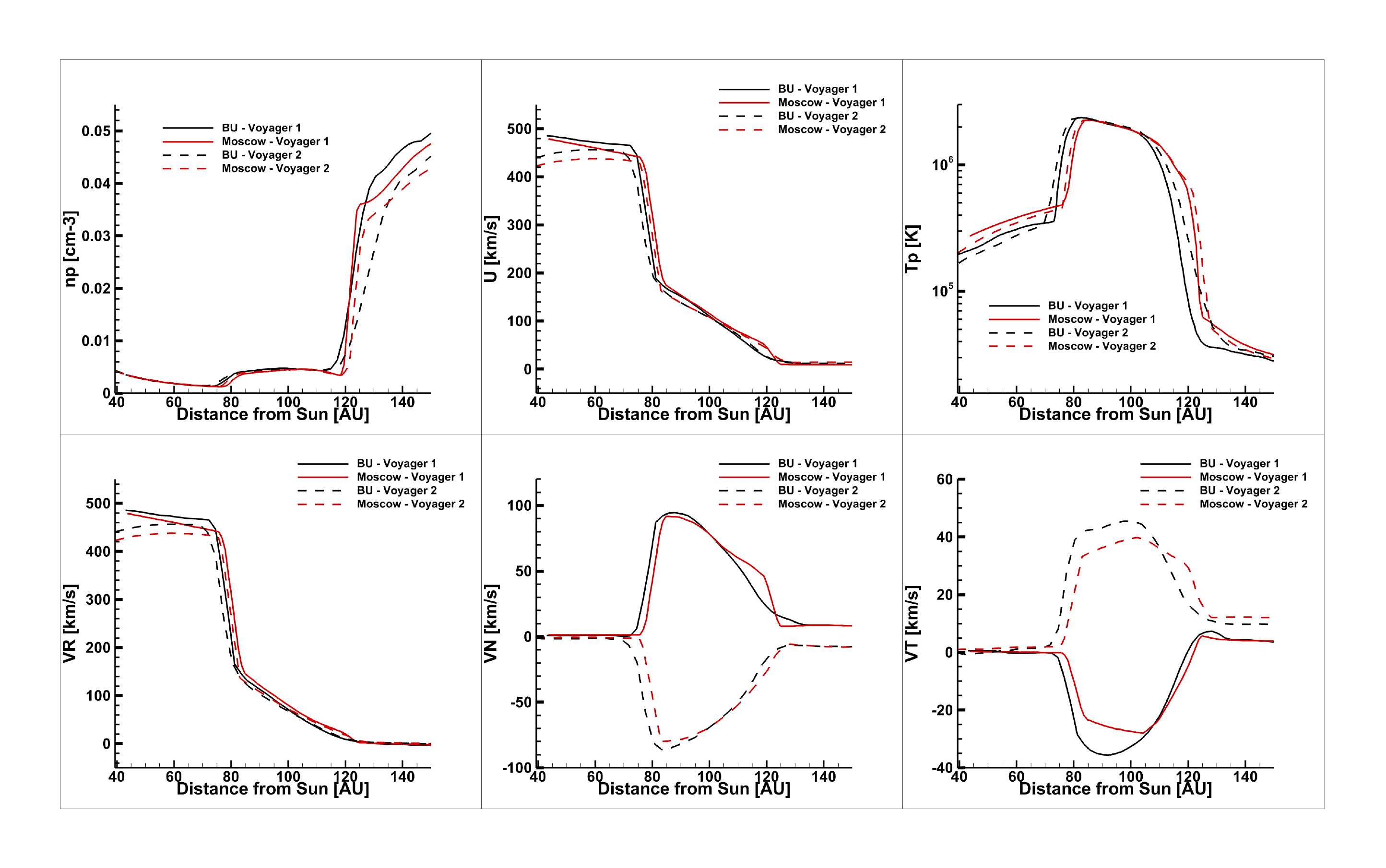}
  \caption{1D cuts along the Voyager 1 (solid) and Voyager 2 (dashed) trajectory for the BU (black) and Moscow (red) models. Top row (from left to right): plasma density, plasma speed, and plasma temperature. Bottom row (from left to right): the radial, normal, and tangential components of the plasma velocity.}
  \label{fig:voyagerplasma}
\end{figure*}

In Figure \ref{fig:voyagerplasma} we present a comparison of the plasma conditions and the magnetic field along the Voyager 1 and 2 trajectories for the two models. While we do not compare with Voyager data directly due to the constant solar wind boundary conditions used in the simulations, the Voyager trajectories are valuable as they are direct probes of heliospheric and ISM conditions by Voyager 1 and 2. Near the poles the heliosheath thickness is notably thicker for the Moscow model (14 AU difference in heliopause location at the northern pole); in contrast the termination shock and heliopause in the models are at similar locations giving similar heliosheath thicknesses (Table \ref{tab:TSHP}). Comparison of these models along the Voyager lines-of-sight reveals interesting characteristics of the heliosphere. The similar results in the Voyager 1 and 2 directions indicate that the shape of the heliotail does not have a significant effect on the heliosheath flows in the direction of the nose. In future work we will compare the BU and Moscow models with time dependent solar cycle conditions corresponding to Voyager observations to compare with the data.

\section{Conclusions}\label{sec:Summary}
Table \ref{tab:summary} summarizes our results. We compared two MHD models using identical inner and outer boundary conditions for the protons and neutral H atoms, and investigated how the different assumptions present in each model influenced the results. Both models produce a split-tail since the solar magnetic field confines the solar particles into distinct north and south columns that become lobes. In the BU model, the ISM flows between the two lobes at large distances. MHD instabilities in the BU model \citep{Opher21}, which are absent in the Moscow model, affect the heliotail and create a short, split tail in contrast with the long tail of the Moscow model. The BU model gives a smaller heliosphere with a heliopause closer to the Sun than predicted in the Moscow model in all directions. The communication between the solar and interstellar magnetic fields at the heliopause in the BU model allows for magnetic reconnection in the port flank \citep{Opher17} and leads to a change in the magnetic topology, which does not occur in the Moscow model.

The BU model uses a polarity for the solar magnetic field that suppresses reconnection at the nose. Reconnection occurs in the port flank, the main location where reconnection should proceed since diamagnetic effects suppress reconnection at the nose \citep{Opher17}. Magnetic reconnection at the heliopause in the BU model twists the interstellar magnetic field and leads to an increase in the interstellar magnetic pressure relative to the Moscow model. 
	
In comparing the two models, we see a short low latitude tail in the BU model and a long tail in the Moscow model. The question of the structure of the heliotail is currently an active and vigorous topic of debate in the community between several groups. This paper establishes that both the BU and Moscow models predict magnetic confinement in the heliotail. This supports the earlier conclusions of the split arrangement of the helitail made by \citet{Opher15}, \citet{Izmodenov15}, and \citet{Pogorelov15}, which had still been debated. The main difference between the two models is the ISM piercing the two lobes in the BU model, which is absent in the Moscow model. The next key step is to pin down the driver for the splitting of the two lobes by the ISM. The relevant driver may well be an instability \citep{Opher21} related to the interaction of streaming ISM neutral H and the ionized heliotail plasma that results in a loss of coherence of the heliospheric jets and leads to the mixing of ISM and  heliospheric tail material. This instability is present in the BU model but not in the Moscow model (for reasons that are not clear at the present time). Investigation of the instability and its consequences such as reconnection and ISM-heliotail plasma mixing are deferred to a future study.

Numerical solution strategies and grid resolutions might play an important role in the different modeling outcomes. This is a topic that needs to be addressed, as global heliosphere modeling suffers from the inability to model phenomena on comparatively small scales, including reconnection, the heliospheric current sheet, and instabilities. Therefore, additional comparisons with other models both on the scale of the global heliosphere and also at small scales are required to further address this.

In future work we will use time dependent MHD models to compare more realistic solutions. The results are important since the different heliosphere properties from these models, such as heliosheath thickness at high latitudes, yield different ENA signatures at different energy bands. These differences are potentially observable by the Interstellar Boundary Explorer. This will be explored in a future study.

\acknowledgments
M.K., M.O., and J.R. were supported by NASA grant 18-DRIVE18$\_$2-0029, Our
Heliospheric Shield, 80NSSC20K0603. The work of I.B., V.I., and D.A. was conducted in the framework of a topic of the state assignment ``Plasma'' to the Space Research Institute, Russian Academy of Sciences. The work at JHU/APL was supported also by NASA under contracts NAS5 97271, NNX07AJ69G, and NNN06AA01C and by subcontract at the Office for Space Research and Technology. Resources supporting this work were provided by the NASA High-End Computing (HEC) program through the NASA 
Advanced Supercomputing (NAS) Division at Ames Research Center. The authors would like to thank the staff at NASA Ames Research Center for the use of the Pleiades 
super-computer under the award SMD-20-46872133.


\newpage

\begin{table}[t]
\centering
\begin{tabular}{ccccccccc}
\tableline
 & & BU (V1) & & Moscow (V1) & & BU (V2) & & Moscow (V2) \\
\tableline
$r_{TS}$ && 82 $\pm$ 4 AU  && 80 $\pm$ 1 AU  && 82 $\pm$ 4 AU  && 80 $\pm$ 1 AU  \\
$r_{HP}$ && 121 $\pm$ 1 AU && 123 $\pm$ 1 AU && 126 $\pm$ 1 AU && 124 $\pm$ 1 AU \\
$d_{HS}$ && 39 AU          && 43 AU          && 44 AU          && 44 AU          \\
\tableline
\end{tabular}
\caption{Comparison of termination shock ($r_{TS}$) and heliopause ($r_{HP}$) locations for the Voyager 1 and 2 directions between the BU and Moscow models, as well as the heliosheath thicknesses ($d_{HS}$). Error bars reflect the grid resolution at the location of the probed boundaries. For the Moscow model, we extract distances using a 1 AU interpolated grid.}
\label{tab:TSHP}
\end{table}

\begin{table}[t]
\tiny
\centering
\begin{tabular}{L|L}
\tableline
How does the size of the heliosphere change between the two models? & The heliopause is closer to the Sun at high latitudes in the BU model than in the Moscow model, leading to a more compressed heliosphere at the poles. \\
\hline
How does the heliotail compare between the two models? & Both models show collimation of the heliosheath plasma by the solar magnetic field. They also both produce a split tail since the solar magnetic field confines the solar particles into two distinct north and south columns that become lobes. In the BU model, reconnection allows the ISM to flow between the two lobes at large distances. The low latitude heliotail is shorter in the BU model while the Moscow model has a long, comet-like tail. \\
\hline
What assumptions have the largest impact on the heliospheric structure? & Magnetic reconnection at the heliopause in the BU model leads to an increase in the magnetic pressure outside of the heliosphere, which compresses the heliosphere. Non-ideal MHD effects contribute to a split tail in the BU model. \\
\hline
How do the plasma properties in the heliosheath compare? & Within 100 AU of the termination shock, the mass flux is similar between the two models, showing an organization by the solar magnetic field. \\
\hline
How do the plasma properties in the ISM compare? & The ISM conditions are similar until roughly 50 AU ahead of the heliopause, where the interaction between the interstellar and solar magnetic field at the heliopause begins to have an effect on the plasma in the BU model. \\
\tableline
\end{tabular}
\caption{Summary of conclusions from this work.}
\label{tab:summary}
\end{table}

\end{document}